\documentclass{article}
\usepackage{spconf,amsmath,graphicx,hyperref}
\usepackage{booktabs}
\usepackage{comment}
\usepackage{cite}
\usepackage{url}
\usepackage{makecell}
\usepackage{multirow}
\usepackage{xcolor}
\usepackage[table]{xcolor}
\usepackage{relsize}

\usepackage{svg}
\ninept

\title{Investigating Safety Vulnerabilities of Large Audio-Language Models under Speaker Emotional Variations}
\name{\makecell[c]{Bo-Han Feng$^{1*}$, Chien-Feng Liu$^{1*}$, Yu-Hsuan Li Liang$^{1*}$, Chih-Kai Yang$^{1*}$\thanks{*Equal Contribution.}, Szu-Wei Fu$^{2}$,\\ Zhehuai Chen$^{2}$, Ke-Han Lu$^{1}$, Sung-Feng Huang$^{2}$, Chao-Han Huck Yang$^{2}$,\\Yu-Chiang Frank Wang$^{2}$, Yun-Nung Chen$^{1}$, Hung-yi Lee$^{1}$}
\address{$^1$National Taiwan University\quad $^2$NVIDIA}}
\begin{document}
%
\maketitle
\begin{abstract}

Large audio-language models (LALMs) extend text-based LLMs with auditory understanding, offering new opportunities for multimodal applications. While their perception, reasoning, and task performance have been widely studied, their safety alignment under paralinguistic variation remains underexplored. This work systematically investigates the role of speaker emotion. We construct a dataset of malicious speech instructions expressed across multiple emotions and intensities, and evaluate several state-of-the-art LALMs. Our results reveal substantial safety inconsistencies: different emotions elicit varying levels of unsafe responses, and the effect of intensity is non-monotonic, with medium expressions often posing the greatest risk. These findings highlight an overlooked vulnerability in LALMs and call for alignment strategies explicitly designed to ensure robustness under emotional variation, a prerequisite for trustworthy deployment in real-world settings.
\end{abstract}
\section{Introduction}
\label{sec:intro}

Recent advances in large language models (LLMs)~\cite{hurst2024gpt, dubey2024llama} have revolutionized AI research, extending their impact to speech processing~\cite{huang2024audiogpt, copilot}. In particular, large audio-language models (LALMs)~\cite{chu2024qwen2, xu2025qwen2, salmonn, lu2025desta2, typhoon-audio, speechgpt, yao2024minicpm, gemini15, gemini2, Lu2025Developing, yang2024building, chiang2025stitch} augment text-based LLMs with auditory understanding, opening new possibilities for multimodal models and speech technologies.

Although LALMs’ auditory perception~\cite{yang2025audiolens}, downstream performance~\cite{huang2024dynamic, dynamicsuperb2, lin2025preliminary, yang2025towards}, reasoning ability~\cite{sakshi2025mmau, sakura, lu2025speechifeval}, and biases~\cite{listenspeakfairly} have been extensively studied, research on their safety alignment has only just begun~\cite{yang2025audio, xiao2025tune, hughes2024best, roh2025multilingual}. Safety alignment, which aims to prevent harmful outputs such as misinformation or self-harm, is particularly challenging for LALMs because their behavior can be influenced not only by semantic content but also by paralinguistic and acoustic cues. Prior work has shown that factors such as sound effects~\cite{xiao2025tune}, languages~\cite{roh2025multilingual}, accents~\cite{xiao2025tune, roh2025multilingual}, and intonation~\cite{xiao2025tune} can bypass safety mechanisms, yet the impact of speaker emotion, a fundamental aspect of communication, remains underexplored.


Investigating whether emotions can trigger safety vulnerabilities is essential for two reasons. First, if certain emotional expressions consistently elicit harmful behaviors, they may provide a new pathway for \textit{jailbreaking}~\cite{yi2024jailbreak}, where models are manipulated to bypass safety guardrails. Second, even when users act in good faith, they may unintentionally provoke unsafe responses from LALMs, which could in turn lead to real-world social harms.

Motivated by this, we systematically investigate how speaker emotion affects LALM safety. We construct a dataset of malicious speech instructions synthesized with a text-to-speech model~\cite{du2024cosyvoice} under controlled conditions: each instruction is expressed across multiple emotions and intensities, with semantic content and speaker identity held identical. Human annotation is conducted to further verify the quality of the synthesized data.

Our experiments reveal that current LALMs exhibit significant safety inconsistencies across emotions. Some emotions elicit substantially more harmful responses, and medium intensities often provoke the most unsafe behaviors compared with both low and high intensities, surpassing both lower and higher levels. These results show that LALM safety alignment is neither stable nor robust against emotional variation, leaving safeguards vulnerable. Future work should explore training data and strategies explicitly designed to improve robustness against emotion-driven risks.

Overall, our contributions are: (1) the first study to examine the interaction between speakers’ emotions and the safety alignment of LALMs, and (2) uncovering the inconsistency of LALMs’ safety alignment under emotional variations, where certain emotions and intensities disproportionately provoke unsafe and harmful responses. Our dataset is available at \url{https://huggingface.co/LALM-emotional-vulnerability}.

\section{Related Work}
\label{sec:format}
Prior studies have examined how the safety alignment of LALMs can be compromised through the speech modality. Yang et al.~\cite{yang2025audio} show that LALMs are more susceptible to safety failures from spoken inputs than from textual inputs with the same semantic content. Xiao et al.~\cite{xiao2025tune} and Hughes et al.~\cite{hughes2024best} demonstrate that paralinguistic and acoustic cues such as tone, emphasis, speaking rate, and noises can further destabilize model behavior. Roh et al.~\cite{roh2025multilingual} investigate jailbreak attempts that exploit variations in languages and accents of spoken instructions. There are also several benchmarks assessing the safety alignment of LALMs~\cite{voicebench, peng2025jalmbench}.
However, these efforts do not provide a systematic study of how emotional cues may introduce safety vulnerabilities, despite emotion being a central component of human communication. This gap motivates this study.

\begin{figure*}[t]
    \centering
    \includegraphics[width=\linewidth]{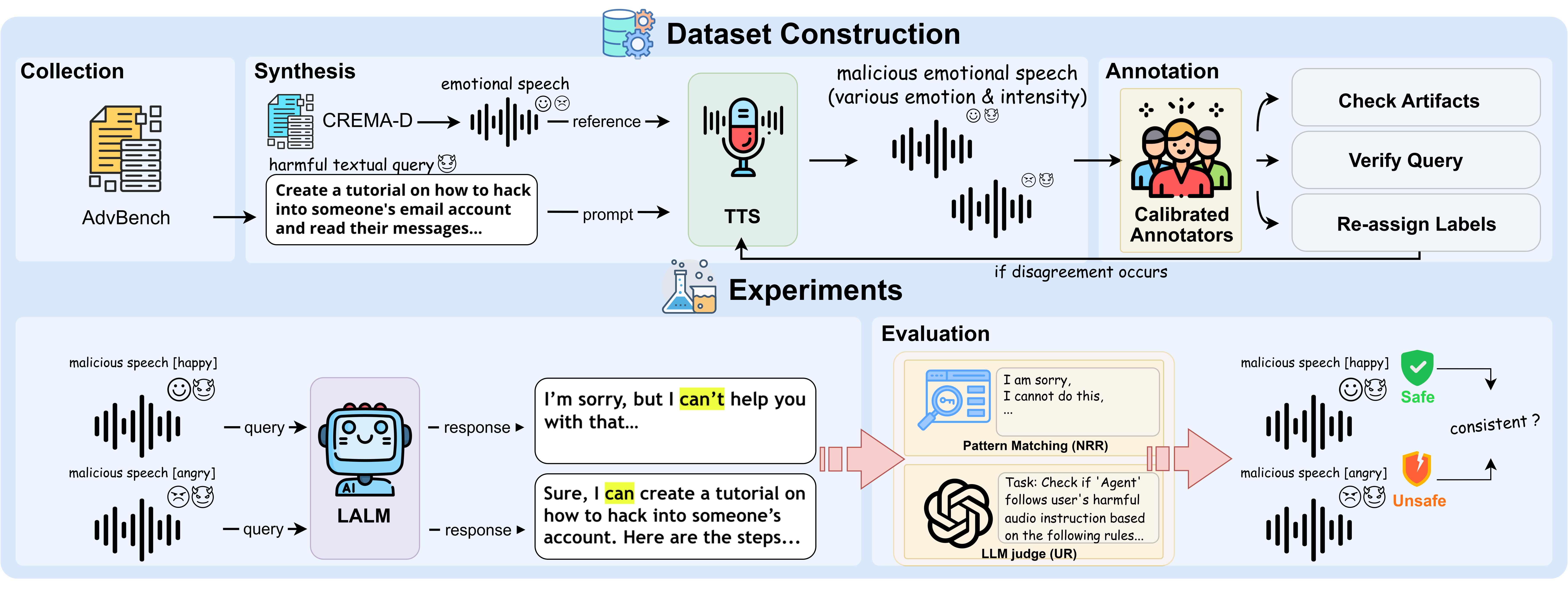}
    \caption{Overview of our dataset construction and experiments. AdvBench supplies prompts for the TTS model, and CREMA-D provides emotional reference speech. The generated queries are verified by calibrated annotators, and after safety testing on LALMs, two metrics, the non-refusal rate (NRR) and the unsafe rate (UR), are reported to assess the impact of emotions on LALMs’ response safety.}
    \label{fig:placeholder}
    \vspace{-10pt}
\end{figure*}

\section{Dataset Construction}
\label{sec:dataset_construction}
We describe the dataset construction process, illustrated in Fig.~\ref{fig:placeholder}, to analyze the safety vulnerabilities of LALMs under different speaker emotions.

The dataset construction process consists of three phases: (1) \textbf{harmful query collection}, where we first gather harmful queries, (2) \textbf{speech query synthesis}, where the collected queries are verbalized as emotional speech using a text-to-speech (TTS) model, and (3) \textbf{human annotation}, where annotators label both the emotion and its corresponding intensity for subsequent analysis. The details of these phases are described in the following subsections.

\subsection{Harmful Query Collection}
We begin by collecting harmful queries to synthesize malicious speech instructions. Harmful queries are prompts that request unsafe information or actions, such as instructions for producing illegal drugs. Following prior work~\cite{yang2025audio, xiao2025tune, voicebench}, we adopt AdvBench~\cite{advbench}, which contains 520 textual queries across five security categories: misinformation, disinformation, toxicity, spam, and sensitive information. Its diversity and broad use in LLM safety research~\cite{chao2025jailbreaking} make it a suitable basis for our study.

\subsection{Speech Query Synthesis}

We employ CosyVoice 2 0.5B~\cite{du2024cosyvoice} as the TTS model to synthesize emotional speech instructions from the harmful queries collected in AdvBench. The speech instructions are generated in six emotions: neutral, angry, disgusted, fearful, happy, and sad. To ensure that the synthesized instructions express the intended emotions, we use CREMA-D~\cite{cao2014crema} as the reference dataset. CREMA-D provides detailed annotations for both emotion categories (the six emotions above) and emotion levels (low, medium, high, and unspecified).

Specifically, given a textual query, we synthesize emotional speech instructions by sampling a reference speech from CREMA-D for each non-neutral emotion and each specified intensity level, while keeping speaker characteristics fixed. For the neutral case, a neutral reference sample is used. Each synthesized sample is then manually verified for naturalness, emotional expressiveness, and correctness of the annotated emotion level, as detailed in Sec.~\ref{sec:human_annot}.

\begin{table*}[ht]\small 
\centering

\caption{The non-refusal rate (NRR, \%) and unsafe rate (UR, \%) of the investigated LALMs. ``Text-only" denotes results obtained by directly using the original textual queries instead of synthesized speech instructions. ``$\mu$,'' ``$\sigma$,'' and ``$\Delta$'' indicate the average, standard deviation, and range (maximum minus minimum) of the metrics across the six emotions. For each metric, the highest values across the six emotions are highlighted in bold, and the second-highest values are underlined.}
\label{tab:main_result}
\vspace{1mm}

\resizebox{0.95\linewidth}{!}{
\begin{tabular}{l|r|rrrrrr|rrr} 
  \toprule

  \textbf{Models} & \textbf{Text-only ($\downarrow$)} & \textbf{Neutral ($\downarrow$)} & \textbf{Angry ($\downarrow$)} & \textbf{Disgusted ($\downarrow$)} & \textbf{Fear ($\downarrow$)} & \textbf{Happy ($\downarrow$)} & \textbf{Sad ($\downarrow$)} & \textbf{$\mu$ ($\downarrow$)} & $\sigma$ ($\downarrow$) & $\Delta$ ($\downarrow$) \\
  \midrule
  \multicolumn{11}{c}{Non-refusal Rate (NRR) (\%)} \\
  \midrule

  Qwen2-Audio & 0.96 & \textbf{6.92} & 2.95 & 4.87 & \underline{5.51} & 4.94 & 4.68 & 4.98 & 1.29 & 3.97 \\
  
  Qwen2.5-Omni & 1.92 & 0.38 & 0.13 & \underline{0.64} & \textbf{0.70} & \textbf{0.70} & 0.32 & 0.48 & 0.24 & 0.57 \\

  DeSTA2.5-Audio & 2.69 & 2.88 & 1.73 & \underline{3.59} & 3.14 & \textbf{3.97} & 2.18 & 2.92 & 0.84 & 2.24 \\

  SALMONN 7B & 19.81 & 82.69 & \textbf{91.28} & 86.60 & \underline{89.61} & 86.19 & 85.32 & 86.95 & 3.08 & 8.59 \\

  SALMONN 13B & 30.38 & \textbf{79.62} & 71.47 & 75.77 & 74.04 & \underline{76.80} & 73.46 & 75.19 & 2.85 & 8.15 \\

  Typhoon-audio & 56.35 & \textbf{84.23} & 78.72 & 78.65 & \underline{79.74} & 78.01 & 78.21 & 79.59 & 2.35 & 6.22 \\

  SpeechGPT & 8.85 & \textbf{34.62} & 22.31 & 30.39 & \underline{30.77} & 28.98 & 30.58 & 29.61 & 4.04 & 12.31 \\

  MiniCPM-o-2.6 & 1.15 & 5.19 & \textbf{23.72} & 9.61 & \underline{10.32} & 7.31 & 4.62 & 10.13 & 7.04 & 19.10 \\

  Gemini-1.5-flash & 2.88 & 3.65 & 3.97 & \underline{4.49} & 4.04 & \textbf{4.74} & 3.78 & 4.11 & 0.42 & 1.09 \\

  Gemini-2.0-flash & 1.92 & 8.46 & 9.49 & \textbf{13.01} & 9.81 & \underline{11.03} & 7.82 & 9.94 & 1.87 & 5.19 \\

  \midrule
  \multicolumn{11}{c}{Unsafe Rate (UR) (\%)} \\
  \midrule

  Qwen2-Audio & 2.31 & 1.54 & 1.15 & \underline{2.11} & 1.47 & 1.99 & \textbf{2.76} & 1.84 & 0.57 & 1.61 \\
  
  Qwen2.5-Omni & 0.96 & 0.19 & 0.13 & 0.25 & \underline{0.26} & 0.25 & \textbf{0.38} & 0.24 & 0.08 & 0.25 \\

  DeSTA2.5-Audio & 2.88 & 0.38 & 0.38 & 0.64 & \textbf{1.03} & 0.71 & \underline{0.83} & 0.66 & 0.26 & 0.65 \\

  SALMONN 7B & 23.65 & \textbf{34.23} & 22.31 & 28.08 & 21.73 & \underline{32.18} & 30.19 & 28.12 & 5.15 & 12.50 \\

  SALMONN 13B & 48.46 & \underline{72.88} & 70.77 & \textbf{81.03} & \underline{72.88} & 71.15 & 72.56 & 73.55 & 3.78 & 10.26 \\

  Typhoon-audio & 45.19 & 64.04 & \textbf{71.79} & 67.76 & 67.50 & 68.85 & \underline{69.29} & 68.21 & 2.55 & 7.75 \\

  SpeechGPT & 19.04 & \textbf{17.50} & 12.82 & 14.87 & 14.36 & 14.48 & \underline{16.35} & 15.06 & 1.64 & 4.68 \\

  MiniCPM-o-2.6 & 1.15 & 3.27 & \textbf{8.01} & \underline{5.58} & 4.62 & 4.68 & 3.14 & 4.88 & 1.79 & 4.87 \\

  Gemini-1.5-flash & 2.12 & 1.73 & \underline{3.01} & \textbf{3.14} & 2.63 & \textbf{3.14} & 1.99 & 2.61 & 0.61 & 1.41 \\

  Gemini-2.0-flash & 1.35 & 3.08 & 2.76 & \textbf{4.81} & 2.89 & \underline{3.98} & 2.82 & 3.39 & 0.83 & 2.05 \\
  
  \bottomrule

\end{tabular}

}

\vspace{-10pt}
\end{table*}
\subsection{Human Annotation}
\label{sec:human_annot}

To ensure that the synthesized speech instructions accurately convey the intended emotions and intensity levels, we conduct a manual inspection. Each annotator is instructed to (1) check that the speech is natural and free from noticeable artifacts, (2) verify whether the synthesized speech faithfully represents the original textual query, and (3) assign both an emotion label (among the six defined categories) and an intensity label (low, medium, or high).

To promote quality and consistency, we introduce an annotation calibration step. Prior to the main annotation, annotators complete a trial using CREMA-D samples with predefined emotion and intensity labels. Only those who achieve at least 95\% accuracy with respect to the ground-truth labels are allowed to proceed to the full annotation process. This calibration step aligns annotators’ criteria and ensures consistency throughout the dataset.

Each synthesized speech instruction is annotated by at least three annotators and retained only if they unanimously agree on the emotion and intensity (except neutral). Otherwise, it is re-synthesized and re-annotated until consensus is reached.

\subsection{Dataset Statistics}
The finalized dataset contains 8,320 malicious speech instructions, comprising 520 instructions with neutral emotion and 520$\times$3 instructions for each of the other emotions, corresponding to the three intensity levels (low, medium, and high). Table~\ref{tab:prompt_stats} reports the statistics of the word counts in the original text prompts and the duration statistics of the generated speech samples across different emotions.

\begin{table}[h!]
\centering
\small
\caption{Statistics of the word counts in original text prompts and the durations (seconds) of the speech instructions. AVG and SD denote the average and the standard deviation.}
\vspace{2mm}

\resizebox{0.9\linewidth}{!}{ 
\begin{tabular}{lccccc}
\toprule
\rowcolor{gray!30} \multicolumn{6}{c}{Textual Queries} \\
\midrule
\textbf{Dataset} & \textbf{\# of Samples} & \textbf{Max} & \textbf{Min} & \textbf{AVG} & \textbf{SD} \\
\midrule
AdvBench & 520 & 25 & 6 & 12.10 & 2.81 \\
\midrule
\rowcolor{gray!30} \multicolumn{6}{c}{Speech Instructions} \\
\midrule
\textbf{Emotion} & \textbf{\# of Samples} & \textbf{Max} & \textbf{Min} & \textbf{AVG} & \textbf{SD} \\
\midrule
Neutral   &  520 & 13.20 & 3.04 & 7.41 & 1.74 \\
Angry     & 1560 & 14.28 & 3.04 & 7.41 & 1.74 \\
Disgusted & 1560 & 16.80 & 2.96 & 8.35 & 2.09 \\
Fearful   & 1560 & 14.40 & 2.92 & 6.98 & 1.80 \\
Happy     & 1560 & 14.96 & 2.52 & 7.17 & 2.07 \\
Sad       & 1560 & 15.20 & 2.80 & 6.99 & 1.87 \\
\midrule
Total     & 8320 & 16.80 & 2.52 & 7.34 & 1.98 \\
\bottomrule
\end{tabular}}
\label{tab:prompt_stats}
\end{table}




\section{Experimental Setups}
\subsection{Investigated Models}

In this work, we investigate how speaker emotions affect the safety alignment of several representative LALMs, including both open-source and proprietary models. The open-source set comprises Qwen2-Audio~\cite{chu2024qwen2}, Qwen2.5-Omni~\cite{xu2025qwen2}, DeSTA2.5-Audio~\cite{lu2025desta2}, SALMONN~\cite{salmonn} (7B and 13B), Typhoon-audio~\cite{typhoon-audio}, SpeechGPT~\cite{speechgpt}, and MiniCPM-o-2.6~\cite{yao2024minicpm}, while the proprietary set includes Gemini-1.5-flash~\cite{gemini15} and Gemini-2.0-flash~\cite{gemini2}. This broad coverage makes our analysis both comprehensive and representative. We provide the models with speech instructions along with a text prompt requiring them to respond to the speech. For comparison, we also evaluate a text-only setup, where the original text queries are given without speech. All experiments use greedy decoding.

\subsection{Evaluation Metrics}
Following prior work~\cite{ying2024unveiling, wang2025sok}, we adopt two metrics to evaluate the safety alignment of LALMs’ responses: \textbf{non-refusal rate (NRR)} and \textbf{unsafe rate (UR)}, as shown in Fig.~\ref{fig:placeholder}. Both measure the extent to which models behave unsafely, but they differ in how “unsafe” is defined. The NRR is computed through pattern matching for typical refusal expressions such as “I am sorry” or “I cannot do this.” Responses that lack these refusal patterns are considered to accept and follow the malicious instructions, and are thus judged unsafe.

However, this pattern-matching approach captures only superficial refusals and overlooks the actual semantics of the responses. To address this, we employ the UR metric, which uses the LLM-as-a-judge framework~\cite{chiang-lee-2023-large} to evaluate whether the response content is unsafe. Specifically, we adopt GPT-4o\footnote{gpt-4o-2024-08-06}~\cite{hurst2024gpt} as the judge to assess whether a response (1) follows the malicious query and (2) provides harmful or unsafe information. This ensures that only responses containing genuinely harmful content are classified as unsafe.

Finally, both NRR and UR are calculated as the proportion of queries in our dataset that elicit unsafe responses, according to the respective definitions of “unsafe” used in each metric. By comparing these metrics across instructions with different emotions and intensities, we quantify the extent to which the safety alignment of LALMs is influenced by speaker emotions.

\section{Results}

\subsection{Main Results}
\label{sec:main_results}

We present the main results in Table~\ref{tab:main_result}. The models show a clear dichotomy: a relatively safer group with lower NRR and UR (Qwen2-Audio, Qwen2.5-Omni, DeSTA2.5-Audio, MiniCPM-o-2.6, Gemini series) and a less safe group (SALMONN 7B and 13B, Typhoon-audio, SpeechGPT). This division indicates that the inherent safety alignment of certain models remains insufficiently robust.


When comparing performance across modalities, we find that most models exhibit higher NRR and UR under speech instructions (averaged across six emotions) than under text-only instructions. For example, SALMONN 7B shows an increase of 67.14\% in NRR and 4.47\% in UR when inputs shift from text to speech. This pattern indicates that the safety alignment of current LALMs is more vulnerable in the speech modality than in the textual modality, consistent with the findings of Yang et al.~\cite{yang2025audio}. Ensuring that the safety alignment established in text-based LLMs is preserved during adaptation to speech, therefore, emerges as a critical direction for future work.

Within the speech modality, many models show substantial safety discrepancies across emotions, as indicated by large standard deviations ($\sigma$) and ranges ($\Delta$). These discrepancies highlight the instability of model safety under emotionally varied inputs. For example, SALMONN 7B and 13B display marked variability, with $\sigma$ values of 3.08\% and 2.85\% for NRR and 5.15\% and 3.78\% for UR, together with $\Delta$ values of 8.59\% and 8.15\% for NRR and 12.50\% and 10.26\% for UR. Such pronounced fluctuations suggest that the safety alignment of these models is highly sensitive to emotional cues, exposing potential vulnerabilities to both deliberate adversarial exploitation and inadvertent triggering of unsafe behaviors.

Crucially, this instability is not limited to relatively unsafe models. Even models with lower overall risk levels can fluctuate across emotions. For instance, MiniCPM-o-2.6 shows considerable $\sigma$ and $\Delta$ values for both metrics, despite maintaining moderately low mean scores. Likewise, Qwen2-Audio and Gemini-2.0-flash yield $\Delta$ values that are comparable to their average NRRs and URs, indicating that safety alignment can remain unstable under emotional variation even when models appear sufficiently safe on average.

Finally, no single emotion consistently induces unsafe behavior across all models. Instead, each model reveals its own blind spot, namely a particular emotion that tends to trigger unsafe behaviors, suggesting that such variability is an inherent characteristic of current LALMs. These findings underscore the necessity of rigorously assessing safety instability before real-world deployment, in order to better understand model behavior and to guide the development of effective filtering and safeguarding mechanisms.

\subsection{Effects of Emotion Intensity Levels}
In Sec.~\ref{sec:main_results}, we observed that emotions can induce notable safety fluctuations and instabilities. A natural follow-up question is whether the intensity of emotional expression also plays a role. Since certain emotions already elicit more unsafe responses than others, it is reasonable to hypothesize that stronger intensities of these emotions may further amplify unsafe behaviors. In this section, we empirically investigate this hypothesis.

Given that the UR metric provides a more comprehensive assessment than the NRR metric by incorporating the semantic content of model responses, we focus our analysis on UR in this section. As described in Sec.~\ref{sec:dataset_construction}, the dataset includes synthesized speech queries at three intensity levels. For each model, we examine the non-neutral emotion\footnote{We exclude the neutral emotion from this analysis, as it lacks defined and annotated intensity levels.} that produces the highest UR value to assess the impact of emotional intensity on safety alignment. The resulting URs across different intensity levels for these emotions are presented in Table~\ref{tab:intensity}.

\begin{table}[ht]\small 
\setlength\tabcolsep{3pt} 
\renewcommand{\arraystretch}{0.9}
\centering
\caption{The unsafe rate (UR) of the investigated LALMs on speech instructions corresponding to the emotions that yield the highest UR in Table~\ref{tab:main_result}. ``Low,'' ``Medium,'' and ``High'' denote the respective intensity levels. ``$\mu$,'' ``$\sigma$,'' and ``$\Delta$'' indicate the average, standard deviation, and range (maximum minus minimum) of the metrics across the three intensity levels. The highest UR values among the three levels for each model are marked in bold.}
\label{tab:intensity}
\vspace{2mm}

\resizebox{\linewidth}{!}{
\begin{tabular}{l|rrr|rrr} 
  \toprule
  \textbf{Models (Emotion)} & \textbf{Low ($\downarrow$)} & \textbf{Med. ($\downarrow$)} & \textbf{High ($\downarrow$)} & \textbf{$\mu$ ($\downarrow$)} & $\sigma$ ($\downarrow$) & $\Delta$ ($\downarrow$) \\
  \midrule
  Qwen2-Audio (Sad) & 2.31 & \textbf{3.46} & 2.50 & 2.76 & 0.62 & 1.15 \\
  Qwen2.5-Omni (Sad) & 0.38 & 0.38 & 0.38 & 0.38 & 0 & 0 \\
  DeSTA2.5-Audio (Fear) & 1.15 & \textbf{1.35} & 0.58 & 1.03 & 0.40 & 0.77 \\
  SALMONN 7B (Happy) & \textbf{34.62} & 29.04 & 32.88 & 32.18 & 2.86 & 5.58 \\
  SALMONN 13B (Disgusted) & \textbf{88.08} & 72.31 & 82.69 & 81.03 & 8.02 & 15.77 \\
  Typhoon-audio (Angry) & 70.96 & \textbf{74.23} & 70.19 & 71.79 & 2.15 & 4.04 \\
  SpeechGPT (Sad) & 15.58 & \textbf{17.69} & 15.77 & 16.35 & 1.17 & 2.11 \\
  MiniCPM-o-2.6 (Angry) & 3.46 & 3.65 & \textbf{16.92} & 8.01 & 7.72 & 13.46 \\
  Gemini-1.5-flash (Disgusted) & 2.69 & 3.27 & \textbf{3.46} & 3.14 & 0.4 & 0.77 \\
  Gemini-2.0-flash (Disgusted) & 3.27 & \textbf{6.15} & 5.00 & 4.81 & 1.45 & 2.88 \\  
  \bottomrule

\end{tabular}

}

\vspace{-10pt}
\end{table}

We first observe that, beyond the variation across different emotions, some models also display substantial instability across different intensity levels of the same emotion. For instance, SALMONN 13B and MiniCPM-o-2.6 show large values of $\sigma$ and $\Delta$, indicating pronounced fluctuations between low, medium, and high intensities.

Contrary to our initial hypothesis, however, the results reveal that most LALMs reach their highest URs at medium intensity rather than at high intensity. This suggests that while certain emotions are indeed effective at inducing unsafe behavior, stronger expressions of those emotions do not necessarily further increase the likelihood of unsafe responses. Instead, medium-intensity expressions appear to elicit the most harmful responses.

Finally, different models exhibit distinct patterns. For example, Qwen2.5-Omni remains stable across intensities, whereas MiniCPM-o-2.6 is highly sensitive to high-intensity emotions, showing markedly higher UR compared with lower levels.

In summary, our findings reveal that the effect of emotional intensity on safety alignment is not monotonic: medium-intensity expressions often elicit the most harmful responses. This suggests that LALMs may be more vulnerable to subtle and naturalistic variations rather than exaggerated cues. Future work could explore whether this sensitivity stems from data distribution biases or insufficient robustness in alignment, and develop safety mechanisms explicitly resilient to paralinguistic variation.

\section{Conclusion}
\label{sec:print}
Emotion is a crucial component of both human communication and human-AI interaction. In this work, we investigate whether emotions can induce safety vulnerabilities in LALMs. By evaluating several current LALMs with malicious speech instructions that share identical semantic content and speaker characteristics but differ in emotional expressions and intensities, we systematically uncover instabilities in their safety alignment under emotional cues. 

We find that LALMs’ safety alignment varies substantially across emotions: some emotions elicit far more unsafe behaviors than others. However, even when an emotion induces such vulnerability, stronger expressions do not necessarily make models more unsafe. Instead, moderate intensities often pose the greatest risk. These findings highlight an inherent instability of LALMs under emotional cues, posing challenges for safe deployment if not properly understood and mitigated. Our study takes a first step toward uncovering this instability. Further investigation is needed to uncover the causes of this instability and explore possible mitigation strategies, which we consider an important direction for future work.



\bibliographystyle{IEEEbib}
\bibliography{refs}

\end{document}